\documentclass[prb,twocolumn,superscriptaddress,showpacs,floatfix]{revtex4}

\usepackage{epsfig,amsmath,amssymb,latexsym}

\usepackage{graphicx}
\usepackage{subfigure}
\begin{document}

\title{Electronic structures of transition metal dipnictides $XPn_2$ ($X$=Ta, Nb; $Pn$=P, As, Sb)} \preprint{1}

 \affiliation{Condensed Matter Group,
  Department of Physics, Hangzhou Normal University, Hangzhou 310036, China}
 \affiliation{Department of Physics, Zhejiang University, Hangzhou 310036, China}

\author{Chenchao Xu}
 \affiliation{Department of Physics, Zhejiang University, Hangzhou 310036, China}
\author{Jia Chen}
 \affiliation{Condensed Matter Group,
  Department of Physics, Hangzhou Normal University, Hangzhou 310036, China}
\author{Guo-Xiang Zhi}
 \affiliation{Department of Physics, Zhejiang University, Hangzhou 310036, China}
\author{Yuke Li}
\author{Jianhui Dai}
 \email[E-mail address: ]{daijh@hznu.edu.cn}
\author{Chao Cao}
 \email[E-mail address: ]{ccao@hznu.edu.cn}
 \affiliation{Condensed Matter Group,
  Department of Physics, Hangzhou Normal University, Hangzhou 310036, China}

\date{\today}

\begin{abstract}
  The electronic structures and topological properties of
transition metal dipnictides $XPn_2$ ($X$=Ta, Nb; $Pn$=P, As, Sb)
have been systematically studied using first-principles
calculations. In addition to small bulk Fermi surfaces, the band
anticrossing features near the Fermi level can be identified from
band structures without spin-orbit coupling, leading to nodal lines
in all these compounds. Inclusion of spin-orbit coupling gaps out
these nodal lines leaving only a pair of disentangled electron/hole
bands crossing the Fermi level. Therefore, the low energy physics
can be in general captured by the corresponding two band model with
several isolated small Fermi pockets. Detailed analysis of the Fermi
surfaces suggests that the arsenides and NbSb$_2$ are nearly
compensated semimetals while the phosphorides and TaSb$_2$ are not.
Based on the calculated band parities, the electron and hole bands
are found to be weakly topological non-trivial giving rise to
surface states. As an example, we presented the
surface-direction-dependent band structure of the surfaces states in
TaSb$_2$.
\end{abstract}

\maketitle

\section{introduction}
Over the last two decades, an exciting development in the condensed
matter physics community has revolutionized the classification of
materials. In addition to the traditional
conductors/insulators/semiconductors, it has been discovered that
symmetries of the crystal may lead to more delicate differences,
giving rise to topologically trivial and non-trivial
phases\cite{TI:3DTI_Kane_Fu,TI:KaneRMP,TI:ZhangRMP}. The difference
between the topological insulators (TI) and normal insulators (NI)
lies at their surfaces, the respective interface with a NI, vacuum.
The topological change across the interface dictates the existence
of a symmetry protected metallic surface state for TI, leading to
distinct magneto-electronic responses. For instance, the metallic
surface state will result in a resistivity plateau in transport
measurement, which is regarded as the hallmark of TI in many
cases\cite{PhysRevB.82.241306,PhysRevB.86.165119,TI:SmB6NM}.

The study of topological materials was taken to a next level with
the discovery of three-dimensional Weyl semimetals. A Weyl semimetal
has separated right- and left-handed nodal points in momentum space
breaking the $\mathcal{T\cdot P}$ symmetry and can thus be realized
either in centrosymmetric topological materials without the
time-reversal $\mathcal{T}$ symmetry
\cite{TI:WeylPr227,TI:WeylHg124}, or in noncentrosymmetric
topological materials without the space inversion or parity symmetry
$\mathcal{P}$ \cite{TI:TaPWeylDFT,TI:TaAsWeylARPES,TI:TaAsFermiArc}.
The presence of the left and right Weyl points will lead to other
intriguing phenomena including formation of disconnected Fermi
arcs\cite{TI:WeylPr227,TI:WeylHg124}. In particular, due to the
chiral anomaly of relativistic Weyl fermions generated by coupling
to the gauge potential, the negative magnetoresistance with unusual
dependence on the electric and magnetic fields may occur in these
materials\citep{TI:WeylNegMR,Weyl:NegMRNbP}.

More recently, a class of transition metal dipnictides has been
experimentally synthesized and
characterized\citep{NbSb2:Petrovic,TaSb2:YKLi,NbTaAs2:TLXia,TaAs2:XDai,TaAs2:YKLuo,TaAs2:SJia}.
All these compounds were reported to exhibit high mobilities and
extremely large positive magnetoresistances (MRs) when the applied
field is perpendicular to the current direction, similar to
previously known Weyl semimetals. Another common feature of these
compounds is the field-induced metal-insulator transition, which in
some cases results in a very clear resistivity plateau at low
temperatures\citep{TaSb2:YKLi,NbTaAs2:TLXia}. Interestingly,
negative longitudinal MRs when the applied field is parallel to the
current direction were reported in TaSb$_2$\citep{TaSb2:YKLi} and
TaAs$_2$\citep{TaAs2:YKLuo}.
All these features imply the rich physics beyond the known TIs and
Weyl semimetals. On the one hand, the large positive MR is mostly
attributed to semiclassical effect of electron-hole
compensation\citep{TaAs2:XDai,TaAs2:SJia}, while the field-induced
metal-insulator transition may be associated with the gap opening in
quasi two-dimensional systems with low carrier density and high
mobility\citep{TaSb2:YKLi}. On the other hand, the field-induced
resistivity plateau and negative MR should have some topological
origins, although their appearances are seemingly
materials-dependent. So far several limited band structure information were occasionally
presented together with the experimental studies\citep{NbSb2:Petrovic,TaSb2:YKLi,TaAs2:XDai,TaAs2:YKLuo}, but the reported
results are far from complete and a thorough comparison for band
structures of all these materials is obviously lacking. Therefore, a
systematic study of electronic structures of this class of materials
is urgently demanded.

In this article, we present our latest first principles results on
these compounds. Our results show a number of common features of
their band structures as well as several important distinctions. In
particular, we shall show that the bulk electronic states of these
materials can be regarded as two-band system with both electron and
hole contributions. The phosphrides (TaP$_2$ and NbP$_2$) are highly
uncompensated with much more holes than electrons, while TaSb$_2$
has slightly more electrons than holes. The rest arsenides and
NbSb$_2$ are nearly compensated. Moreover, both the electron and
hole bands are weakly topological in all these materials, leading to
protected surface states of these compounds.

\section{method}
The calculations were performed with density functional theory (DFT)
as implemented in Vienna Abinitio Simulation Package
(VASP)\cite{method:vasp,method:pawvasp}. Plane-wave basis up to 400
eV were employed in the calculations. Throughout the calculation, the 
PBE parameterization of generalized gradient approximation to the exchange 
correlation functional was used\cite{method:pbe}. The crystal structure was
fully optimized using the conventional cell with $4\times 12\times
5$ $\Gamma$-centered K-mesh until the force on each atom less than 1
meV/\AA\ and internal stress less than 0.1 kbar. The subsequent
electronic structure calculations were then performed with primitive
cell and a $8\times 8\times 5$ $\Gamma$-centered K-mesh.

The topological indices $\mathcal{Z}_2$ were calculated using the
parity-check method proposed by Fu {\it et al.}\cite{TI:z2inv}. The
DFT band structures were fitted to a tight-binding (TB) model
Hamiltonian using the maximally localized wannier function (MLWF)
method\cite{method:mlwf} with Ta-5d and Sb-5p orbitals. The Fermi
surfaces were then obtained by extrapolating the TB hamiltonian on a
$100\times 100\times 100$ K-mesh; and the surface states were
calculated using the TB hamiltonian by calculating the surface
Green's function\cite{method:surfgf}.

\section{results and discussion}
  \subsection{Crystal Structure and Brillouin Zone}
\begin{figure}
 \subfigure[]{
   \includegraphics[width=3.6 cm]{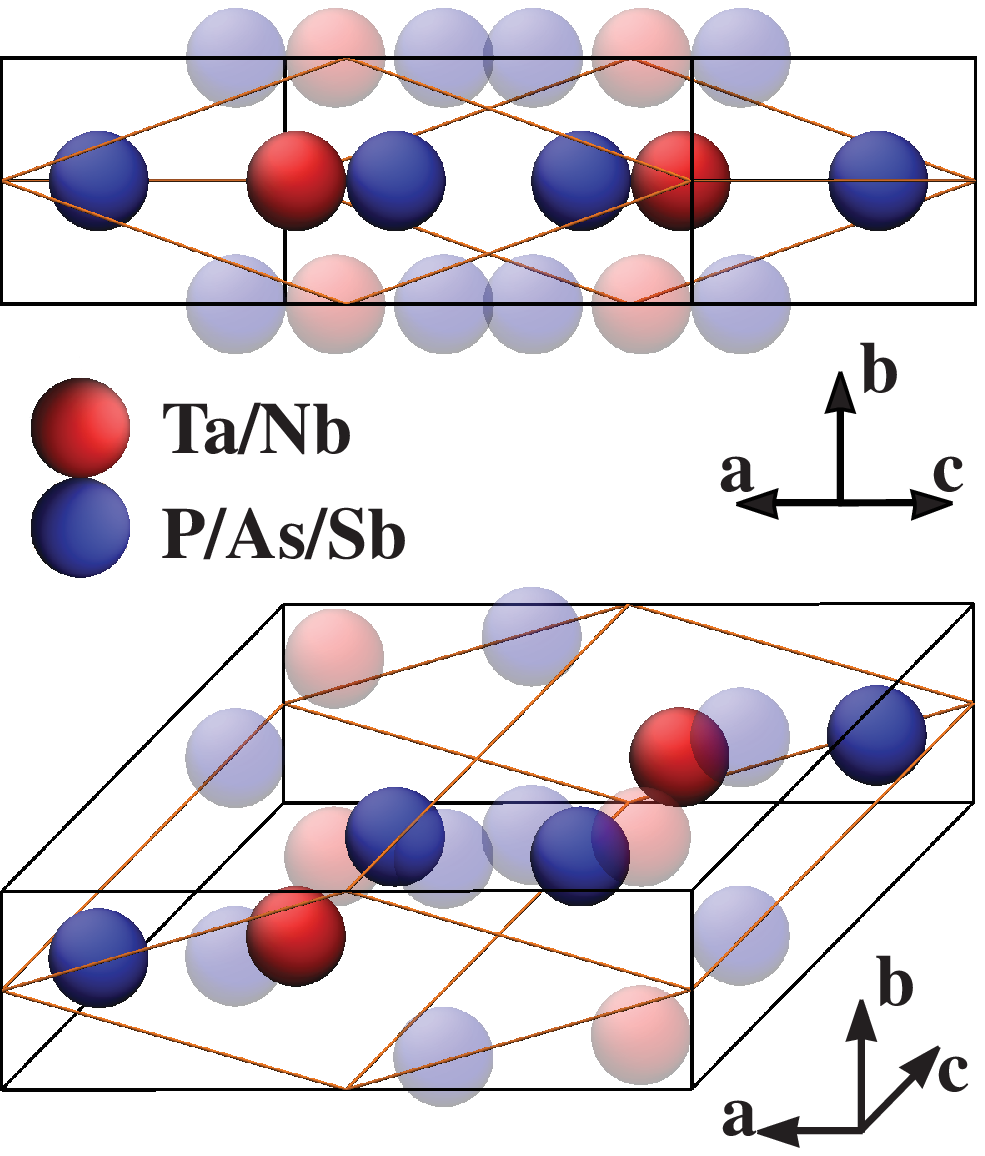}}
 \subfigure[]{
   \includegraphics[width=4.4 cm]{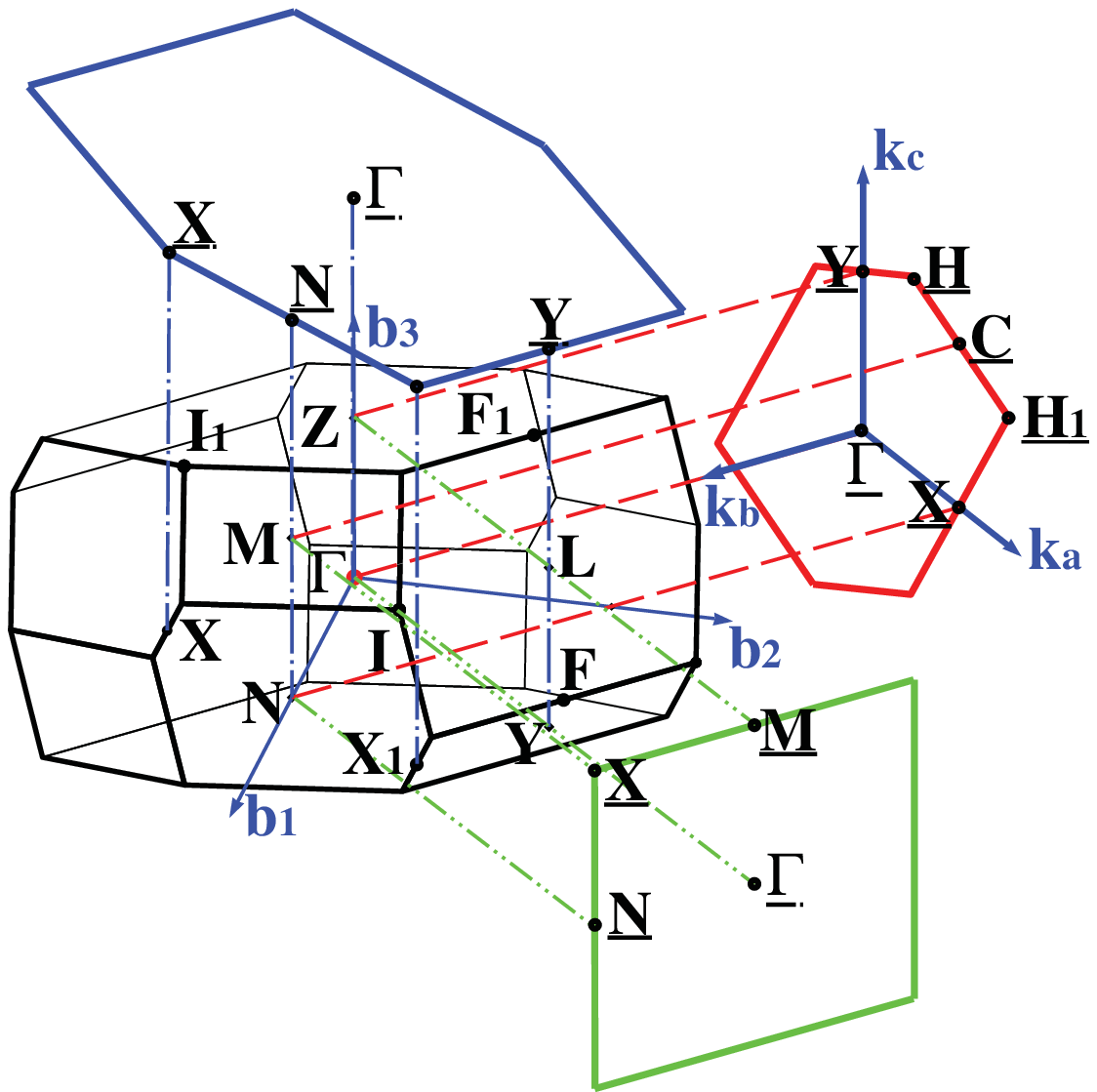}}
\caption{(a) Conventional (the larger black box) and primitive (the smaller orange box) cells of $XPn_2$. Transparent atoms only appears in the conventional cell. (b) Primitive Brillouin zone (BZ), the reciprocal lattice vectors of conventional BZ ($k_a$, $k_b$, and $k_c$); the BZ of conventional [100] (green solid lines), [010] surface (red solid lines), [001] surface (blue solid lines), and their respective high symmetry points. \label{fig:geometry}}
\end{figure}

First of all, we show the optimized geometry parameters of $XPn_2$
compounds with spin-orbit coupling (SOC) effect considered (TABLE
\ref{tab:geometry})\footnote{The geometry parameters obtained 
without SOC differs by less than 1\% compared to those obtained with 
SOC.}. All compounds share the same centrosymmetric
base-centered monoclinic structure with space group $C_{12/m1}$, as
shown in Figure \ref{fig:geometry}(a). The calculated lattice 
constants as well as the atomic coordinates are
within 5\% errorbar compared to the experimental values, manifesting
the validity of our calculations. Since the conventional unit cell
consists of 2 primitive cells, the primitive Brillouine zone (BZ) is
twice large as the conventional BZ. The BZ for the primitive cell,
as well as the definition of high symmetry points, is illustrated in
Fig. \ref{fig:geometry}(b).

\begin{table*}
 \caption{Optimized geometry parameters of $XPn_2$ compounds. $\beta$ is the angle formed by $\mathbf{a}$ and $\mathbf{c}$ lattice vectors. The numbers in the parenthesis are experimental values for comparison purposes, from Ref. \cite{TaNbP2},\citep{NbTaAs2:TLXia},\citep{NbSb2:Petrovic},\cite{TaNbP2},\citep{TaAs2:XDai},\citep{TaSb2:YKLi} for NbP$_2$, NbAs$_2$, NbSb$_2$, TaP$_2$, TaAs$_2$ and TaSb$_2$, respectively; whereas the internal coordinates for antimonides and TaAs$_2$ are from Ref. \citep{FURUSETH:1964aa} and \citep{TaAs2:YKLuo}, respectively. \label{tab:geometry}}
 \begin{tabular}{c|c|c|c|c|c|c}
  \hline
    & NbP$_2$\cite{TaNbP2} & NbAs$_2$\citep{NbTaAs2:TLXia} & NbSb$_2$\citep{NbSb2:Petrovic} & TaP$_2$\cite{TaNbP2} & TaAs$_2$\citep{TaAs2:XDai} & TaSb$_2$\citep{TaSb2:YKLi} \\
  \hline
  a (\AA) & 8.902 (8.872) & 9.454 (9.354) & 10.359 (10.233) & 8.892 (8.861) & 9.441 (9.329) & 10.356 (10.223) \\
  b (\AA) & 3.290 (3.266) & 3.418 (3.381) &  3.676 (3.630) & 3.290 (3.268) & 3.422 (3.385) &  3.697 (3.645) \\
  c (\AA) & 7.584 (7.510) & 7.884 (7.795) &  8.423 (8.328) & 7.543 (7.488) & 7.843 (7.753) &  8.385 (8.292) \\
  $\beta$ & 118.97 (119.10) & 119.40 & 120.04 (120.04) & 119.23 (119.31) & 119.70 (119.70) & 120.52 (120.39) \\
  $x_{X}$ & 0.1598 (0.154) & 0.1569 (0.154) & 0.1516 (0.157) & 0.1601 (0.154) & 0.1569 (0.157) & 0.1504 (0.16) \\
  $z_{X}$ & 0.2007 (0.200) & 0.1961 (0.196) & 0.1902 (0.196) & 0.2002 (0.200) & 0.1954 (0.196) & 0.1883 (0.20)\\
$x_{Pn,I}$ & 0.4029 (0.399) & 0.4050 (0.399) & 0.4050 (0.404) & 0.4037 (0.399) & 0.4055 (0.406) & 0.4051 (0.40)\\
$z_{Pn,I}$ & 0.1000 (0.112) & 0.1065 (0.107) & 0.1116 (0.196) & 0.1012 (0.112) & 0.1075 (0.107) & 0.1125 (0.11)\\
$x_{Pn,II}$ & 0.1334 (0.143) & 0.1414 (0.140) & 0.1489 (0.142) & 0.1331 (0.143) & 0.1406 (0.139) & 0.1486 (0.14)\\
$z_{Pn,II}$ & 0.5258 (0.531) & 0.5286 (0.526) & 0.5358 (0.527) & 0.5251 (0.531) & 0.5275 (0.526) & 0.5351 (0.53)\\
  \hline
 \end{tabular}
\end{table*}

\subsection{Bulk band structure and DOS}

\begin{figure*}
 \includegraphics[width=16cm]{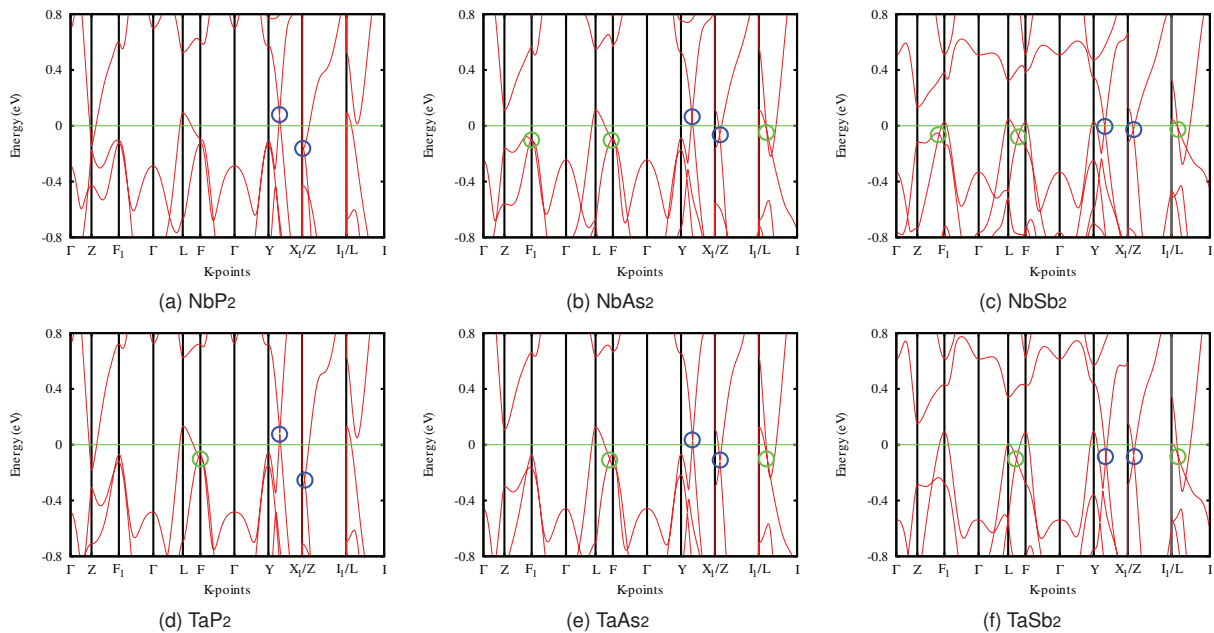}
\caption{Bulk band structure of $XPn_2$ calculated without SOC (reproduced with MLWF fitted TB Hamiltonians). The circles indicates anticrossing features within $E_F\pm$200 meV range. The Fermi energy is aligned at 0. The blue circled features are present in all compounds, while the green circled features are not. \label{fig:bulkbs}}
\end{figure*}

With the optimized structure, we calculated the bulk band structures
without the SOC effect (FIG. \ref{fig:bulkbs}). The overall band
structures of these compounds resemble each other, while the
systematic changes can be traced. For all the compounds, there are
one electron band and one hole band which cross the Fermi level
$E_F$, suggesting the two-band feature and coexistence of
electron/hole carriers in these materials. In addition, two
anti-crossing features can always be identified within 0.2 eV range of
the Fermi level $E_F$ as indicated by the blue circles. These band
inversion properties can be verified with symmetry analysis for the
states near the K-points. By explicitly calculating the coordinates
of the nodal k-points for the TaSb$_2$ compound, we have found that 
these band crossing k-points form nodal lines, consistent with the 
general argument posed by Weng {\it et al.}\cite{TI:NLS_Graphene,TI:NLS_Cu3PdN}. For phosphorides, 
the two anti-crossings are energetically separated and far from $E_F$. 
They are closer in arsenides, and eventually become energetically almost
degenerate and very close to $E_F$ in antimonides. Furthermore, the
direct gap at Z from the electron to hole band is increased from a
few meVs in NbP$_2$ to $\sim$ 400 meV TaSb$_2$, reflecting the
increase of atom sizes and the bond lengths. Similar changes can
also be seen around F, F$_1$, and in between L and I. Thus,
more anti-crossing features can be identified in either arsenides
or antimonides, as indicated by the green circles in FIG.
\ref{fig:bulkbs} (b-c) and (e-f). It is also interesting to notice
that the anti-crossing features from L to I are highly asymmetric when
present, different from all others.

\begin{figure*}
 \includegraphics[width=16cm]{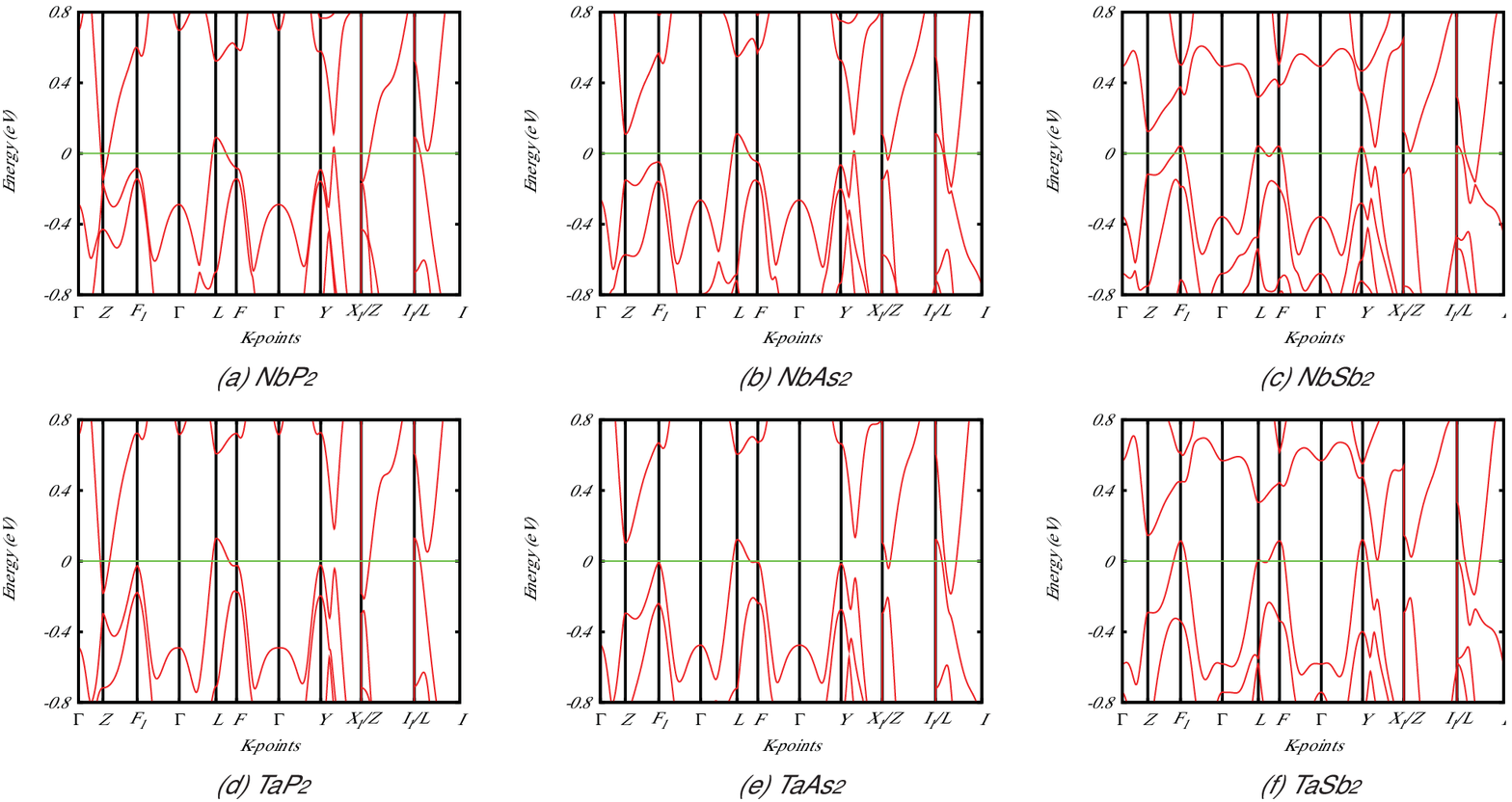}
\caption{Bulk band structure of $XPn_2$ calculated with SOC. The Fermi energy is aligned at 0.\label{fig:bssoc}}
\end{figure*}

\begin{figure}
 \includegraphics[width=8cm]{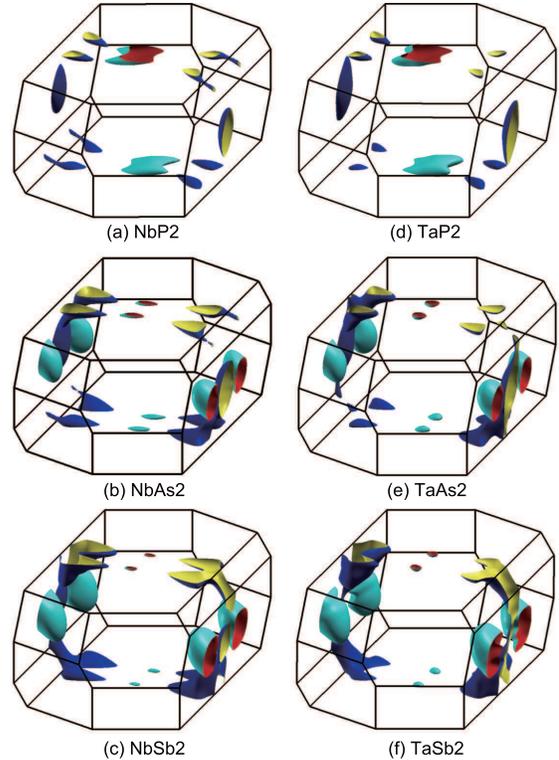}
\caption{Fermi surfaces of $XPn_2$ calculated with SOC (using MLWF fitted TB Hamiltonians). In all compounds, there are 3 isolated pockets, including 2 large ones and 1 small one. \label{fig:fermisurf}}
\end{figure}

Once the SOC effect is taken into consideration (FIG. \ref{fig:bssoc}), all the above nodal lines become gapped, leaving us a pair of fully gapped hole band and electron band.  Furthermore, these bands are also completely separated from all other bands in the whole BZ. Therefore, $\mathcal{Z}_2$ invariants can be calculated for each of these bands. Apart from the gap of nodal lines, some other interesting features can be identified in FIG. \ref{fig:bssoc}. Firstly, the size of the SOC splitting roughly follows the order of Nb$<$Ta and P$<$As$<$Sb, in consistent with normal expectation. Secondly, the gapped states close to Z forms two very small electron pockets for arsenides and antimonides, while the gapped states close to Y forms two small hole pockets for NbP$_2$ and NbAs$_2$. The formation of these small closed pockets can also be evidenced from the Fermi surface (FS) plot of these compounds (FIG. \ref{fig:fermisurf}). The DOS at the Fermi level $n(E_F)$ are evaluated to be 0.099 (NbP$_2$), 0.546 (NbAs$_2$), 1.084 (NbSb$_2$), 0.115 (TaP$_2$), 0.583 (TaAs$_2$), and 0.803 (TaSb$_2$) states per unit cell per eV, respectively. The systematic increase of $n(E_F)$ from phosphrides to antimonides is consistent with the increase of FS area; and proportional to the carrier density in these compounds and the metallicity thereafter.

Closer examination of the FS plots reveals further details of these compounds. From phosphorides to antimonides, the increase of $n(E_F)$ is due to increase of both hole FS (blue/green sheets) and electron FS (cyan/red sheets) (FIG.\ref{fig:fermisurf}). We can also identify the electron/hole DOS contribution by evaluating the second derivative of band energy $m_{\alpha\beta}=d^2\epsilon_\mathbf{k}/dk_{\alpha}dk_{\beta}$ for all bands crossing the Fermi level. It has been proposed that the large MR observed was due to nearly compensated electron/hole density\citep{TaAs2:XDai,NbTaAs2:TLXia}. As the current direction is usually aligned along $\mathbf{b}$ axis, we classified the contribution according to $m_{bb}$. The electron/hole DOS ratio for these compounds are 1:2.4 (NbP$_2$), 1:1.13 (NbAs$_2$), 1.05:1 (NbSb$_2$), 1:1.9 (TaP$_2$), 1:1.16(TaAs$_2$), and 1.25:1 (TaSb$_2$), respectively. Thus, the phosphrides are highly uncompensated with much more holes than electrons, and TaSb$_2$ has slightly more electrons than holes, while all other compounds are nearly compensated. Although such classification is very crude indeed, it agrees with previous experimental results\citep{NbTaAs2:TLXia,TaAs2:SJia,TaAs2:XDai,TaSb2:YKLi}.

 \subsection{$\mathcal{Z}_2$ invariants}
\begin{table}
 \caption{Parities of bands at time-reversal invariant momenta (TRIM) $\Gamma$: (0,0,0), N: ($\pi$, 0, 0), N$^{\prime}$: (0, $\pi$, 0), Y: ($\pi$, $\pi$, 0), Z: (0, 0, $\pi$), M: ($\pi$, 0, $\pi$), M$^{\prime}$: (0, $\pi$, $\pi$) and L: ($\pi$, $\pi$, $\pi$). $\Pi_n$ is the multiplication of the parities for bands 1 to $n$. The highest occupied band at each TRIM is indicated with $^\circ$. \label{parity}}
 \begin{tabular}{c||c|c|c|c||c|c|c|c||c|c|c|c}
  \hline
  & \multicolumn{4}{c||}{$X$P$_2$} & \multicolumn{4}{c||}{$X$As$_2$} & \multicolumn{4}{c}{$X$Sb$_2$} \\
  \hline
   & $\Pi_{20}$ & $\Pi_{21}$ & $\Pi_{22}$ & $\Pi_{23}$ & $\Pi_{20}$ & $\Pi_{21}$ & $\Pi_{22}$ & $\Pi_{23}$ & $\Pi_{20}$ & $\Pi_{21}$ & $\Pi_{22}$ & $\Pi_{23}$ \\
  \hline
  $\Gamma$      & $+$ & $-^\circ$ & $+$ & $-$   & $+$ & $-^\circ$ & $+$ & $-$   & $+$ & $-^\circ$ & $+$ & $-$ \\
  N             & $+$ & $+^\circ$ & $+$ & $+$   & $+$ & $+^\circ$ & $+$ & $-$   & $+$ & $+^\circ$ & $+$ & $-$ \\
  N$^{\prime}$  & $+$ & $+^\circ$ & $+$ & $+$   & $+$ & $+^\circ$ & $+$ & $-$   & $+$ & $+^\circ$ & $+$ & $-$ \\
  Y             & $+$ & $+^\circ$ & $+$ & $-$   & $+$ & $+^\circ$ & $+$ & $-$   & $+^\circ$ & $+$ & $+$ & $-$ \\
  Z             & $+$ & $-$ & $-^\circ$ & $+$   & $+$ & $-^\circ$ & $-$ & $+$   & $+$ & $-^\circ$ & $-$ & $+$ \\
  M             & $+$ & $-^\circ$ & $-$ & $-$   & $+$ & $-^\circ$ & $-$ & $-$   & $+$ & $-^\circ$ & $-$ & $-$ \\
  M$^{\prime}$  & $+$ & $-^\circ$ & $-$ & $-$   & $+$ & $-^\circ$ & $-$ & $-$   & $+$ & $-^\circ$ & $-$ & $-$ \\
  L             & $+^\circ$ & $+$ & $-$ & $+$   & $+^\circ$ & $+$ & $-$ & $-$   & $+^\circ$ & $+$ & $-$ & $-$ \\
  \hline
 \end{tabular}
\end{table}

Since the structure has an inversion center, the topological indices
can be easily calculated by examining the parities of occupied bands
at the time-reversal invariant momenta (TRIM)\cite{TI:z2inv}.
Noticing that both the 21st and 22nd bands cross the Fermi level
$E_F$, we calculated four sets of $\mathcal{Z}_2$ numbers $\Pi_{20}$,
$\Pi_{21}$, $\Pi_{22}$ and $\Pi_{23}$, where $\Pi_n$ is the
multiplication of the parities for bands 1 to $n$. The $\Pi_{20}$,
$\Pi_{21}$ and $\Pi_{22}$ are exactly the same for all these
compounds, and $\Pi_{20}$ for all TRIMs are +1, suggesting that the
fully occupied bands are topologically trivial. Although the
occupations at each TRIM vary with respect to the pnictogen, the
$\mathcal{Z}_2$ classification can be determined to be $(0;111)$ for all these
compounds. Nevertheless, the emergence of surface state of weakly
topological material is not universal and depends on the details of
TRIMs of the specific surface. We shall clarify this in the next
section. Interestingly, $\Pi_{23}$s of both arsenides and
antimonides correspond to $\mathcal{Z}_2$ classification $(1;111)$, and is
topologically strong. This may lead to new topological insulators in
similar compounds.

\subsection{Surface states}

\begin{figure}
 \includegraphics[width=8cm]{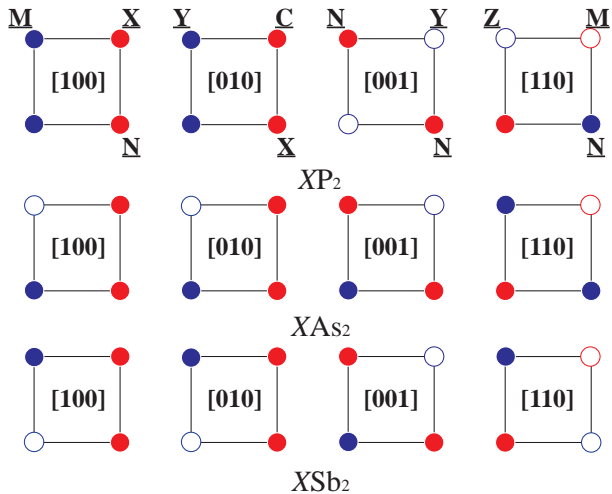}
\caption{Cartoon illustration of possible surface states. The rows from top to bottom are phosphorides, arsenides, and antimonides, respectively. The bottom left corner of each panel is \underline{$\Gamma$}, and the surface indices are illustrated at the center of the panel. The blue/red circles represent $+$/$-$ polarizations, respectively; and open circles indicate overwhelming bulk states due to occupations. Topological surface states can be present between solid circles with different colors. \label{fig:surfstate}}
\end{figure}

In order to calculate the surface state in these compounds, we first
identify the 2D BZs for certain important surfaces. Since the
conventional cell is constructed from the primitive cell using
$\mathbf{a}=\mathbf{a'}+\mathbf{b'}$,
$\mathbf{b}=-\mathbf{a'}+\mathbf{b'}$ and $\mathbf{c}=\mathbf{c'}$;
the [100] surface of the primitive cell is therefore [110] surface
of the conventional cell. To avoid confusion, we shall label the
surfaces using conventional cell index from now on. The 2D BZs of
the [100], [010], and [001] surfaces are illustrated in the figure
\ref{fig:geometry}(b).

\begin{figure}
 \includegraphics[width=8cm]{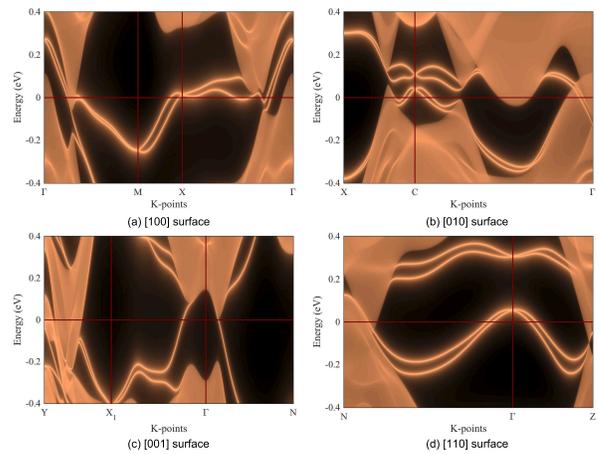}
\caption{Surface states of TaSb$_2$. The surface direction is labeled in conventional cell, thus [110] surface is indeed [100] surface of the primitive cell. \label{fig:surfbs}}
\end{figure}

As we have identified previously, the electron polarizations at each
TRIM are exactly the same for all these materials, in spite of the
different occupations involved. Therefore, the presence of the
metallic surface state depends on two factors: the polarization at
the TRIMs of the specific surface, and the occupations at those
TRIMs. The latter is not relevant in insulators since the universal
bulk gap ensures the occupations at all TRIMs to be the same. It is
not the case in topological metals because it may introduce metallic
bulk state. Indeed, there are 42 valence electrons in the primitive 
cell, corresponding to 21 filled bands if the compound is an insulator. 
Therefore, if any of the two bulk TRIM occupations corresponding to
the same surface TRIM is different from 21, bulk state would
overwhelm around the specific surface TRIM, suppressing the
emergence of the surface state near it. Away from this TRIM,
metallic surface states can still emerge, but will not be symmetry
protected. With above statement in mind, we can identify symmetry
protected surface states in these materials as illustrated in FIG.
\ref{fig:surfstate}. The surface states can be present between two
TRIMs only if both TRIMs are solid and filled with opposite colors. We demonstrate the
surface states of TaSb$_2$ in detail in FIG. \ref{fig:surfbs}.

\section{conclusion}

In conclusion, we have performed a systematic study of the
electronic structures and topological properties of transition metal
dipnictides $XPn_2$ ($X$=Ta, Nb; $Pn$=P, As, Sb) using
first-principles calculations. Nodal line features can be identified
in these compounds, and the inclusion of spin-orbit coupling gaps
out all the anticrossing features. Small electron (hole) pockets due
to the gapped nodal states can be identified in arsenides and
antimonides (NbP$_2$ and NbAs$_2$), respectively. The DOS at $E_F$
systematically increases in the order of phosphorides $<$ arsenides
$<$ antimonides. Furthermore, the NbAs$_2$, NbSb$_2$ and TaAs$_2$
are nearly compensated semimetals; the phosphorides have much more
holes than electrons; while the TaSb$_2$ has slightly more electrons
than holes. By calculating the band parities, we found both the
electron and hole bands are weakly topological and thus shall give
rise to surfaces states, although the presence of these states are
not as robust as the ones emerging from strongly topological
insulators and depends on details including the electron
polarizations at TRIMs and the electron occupations.

\begin{acknowledgments}
This work has been supported by the 973 project (No. 2014CB648400), the
NSFC (No. 11274006, No. 11474082 and No. 11274267) and the NSF of Zhejiang
Province (No. LR12A04003 and No. LZ13A040001). All calculations were performed at the
High Performance Computing Center of Hangzhou Normal University
College of Science.
\end{acknowledgments}

Note Added: Before the submission of this manuscript, we became aware of the work by Shen {\it et al.}\cite{NbAs2:NegMR}. They have also observed negative MR in NbAs$_2$, and their band structure results are in good agreement with ours. We notice that the three Fermi pockets in their calculation is due to the folded BZ for the conventional cell instead of the primitive one. The $F_c$ frequency from their experiment may be due to the two small pockets close to the Z point in our calculation which become indistinguishable because of the identical shape and small sizes.

\bibliography{XSb2}

\end{document}